\documentclass{PoS}
\usepackage[percent]{overpic}

\newcommand{\fermi}{{\it Fermi}}
\newcommand{\fermilat}{{\it Fermi}--LAT}
\newcommand{\gray}{$\gamma$-ray}

\newcommand{\apj}{{ApJ}}
\newcommand{\apjs}{{ApJS}}
\newcommand{\aap}{{A\&A}}
\newcommand{\aj}{{Astron.\ J.}}
\newcommand{\nat}{{Nature}}
\newcommand{\araa}{{Ann.\ Rev.\ Astron.\ \& Astrophys.}}
\newcommand{\hess}{{H.E.S.S.}}
\newcommand{\GP}{{\it GALPROP}}

\newcommand{\hi}{H~{\sc i}}

\title{Observations of High-Energy Gamma-Ray Emission Toward the Galactic Centre with the \fermi\ Large Area Telescope}

\ShortTitle{Observations of High-Energy \gray{s} Toward the Galactic Centre with the \fermilat}

\author{{\speaker{Troy~A.~Porter}$^a$ and Simona Murgia$^b$ on behalf of the \textit{Fermi}-LAT Collaboration}\\
        \llap{$^a$}Hansen Experimental Physics Laboratory and Kavli Institute for Particle Astrophysics and Cosmology,
        Stanford University, Stanford, CA 94305, U.S.A.\\
        \llap{$^b$}Department of Physics and Astronomy, University of California, Irvine, CA 92697, U.S.A.\\
        E-mail: \email{tporter@stanford.edu}, \email{smurgia@uci.edu}}

\abstract{The inner region of the Milky Way is one of the most interesting 
and complex regions of the \gray{} sky. 
The intense interstellar emission and resolved point sources, as well as 
potential contributions by other sources such as unresolved source 
populations and dark matter, complicate the interpretation of the data. 
In this contribution the \fermilat\ team analysis of a 
$15^\circ \times 15^\circ$ region about the Galactic centre is described.
The methodology for point-source detection and treatment 
of the interstellar emission is given. 
In general, the bulk of the \gray{} emission from this region 
is attributable to a combination of these two contributions. 
However, low-intensity residual emission remains and its 
characterisation is discussed.}

\FullConference{The 34th International Cosmic Ray Conference,\\
		30 July- 6 August, 2015\\
		The Hague, The Netherlands}

\begin{document}

\section{Introduction}
The region surrounding the Galactic centre (GC) is among the brightest and 
most complex in high-energy \gray{s}, with on-going massive star 
formation providing all types of known or suspected 
cosmic ray (CR) and \gray{} sources. 
The GC also houses a $\sim 10^6$ M$_\odot$ black hole (e.g., \cite{2010RvMP...82.3121G}) 
and the region is predicted to 
be the brightest source of \gray{s} associated with 
annihilation or decay of massive weakly-interacting particles 
(see the reviews by, e.g., \cite{1996PhR...267..195J, 2000RPPh...63..793B, 2010ARA&A..48..495F}).  
Despite detection in the 100~MeV to GeV range by the EGRET instrument 
on the {\it Compton Gamma-Ray Observatory} \cite{EGRETGCref} and 
at higher energies by the \hess\ Cherenkov array 
\cite{2006Natur.439..695A,2006ApJ...636..777A} the 
characterisation of the \gray{} emission for $<100$~GeV energies 
in the region surrounding the GC has remained elusive.

The \gray{} emission in the Galaxy is predominantly due to the 
interactions of CR particles with the interstellar gas and 
radiation fields.
This interstellar emission is a fore-/background against which \gray{} 
point sources are detected.
In the Galactic plane, and particularly toward the GC, 
the intensity of this emission makes disentangling 
the contributions by \gray{} point sources and truly diffuse processes 
particularly challenging.

Since 2008 the Large Area Telescope instrument on the \fermi{} Gamma-Ray 
Space Telescope (\fermilat) has been taking data in the range 20~MeV to more 
than 300~GeV energies.
In this contribution, an analysis 
is described of the \gray{} emission observed by 
the \fermilat\ during the first 62~months of the mission toward the inner 
Milky Way that characterises 
the $15^\circ\times15^\circ$ region in Galactic coordinates centred on the GC.
This encompasses the innermost $\sim 1$~kpc where
the CR intensities, interstellar gas and radiation field densities are 
highest but most uncertain, and signatures of new physics may be detectable.
The analysis uses multiple interstellar emission models (IEMs) 
together with an iterative fitting procedure 
to self-consistently determine the contributions by diffuse and 
discrete sources of high-energy \gray{} emission. 
The \GP\ CR propagation code\footnote{For a detailed description of the \GP\ 
code the reader is referred to the dedicated website: http://galprop.stanford.edu}~(e.g., \cite{1998ApJ...493..694M,2012ApJ...752...68V}) 
is used to calculate components of IEMs that are fit to the \fermilat\ data to 
predict the 
interstellar emission fore-/background toward the $15^\circ\times15^\circ$ 
region.
Candidate locations of point sources are found using a 
wavelet-based algorithm \cite{1997ApJ...483..350D,Ciprini:2007zz}.
These are used together 
with the IEMs to define a model for the emission of the region, 
which is then optimised in a maximum-likelihood fit to determine
the contribution by CR-induced diffuse emission from the innermost 
$\sim 1$~kpc and \gray{} point sources.

\section{Methodology}

\subsection{Data Selection and Preparation}
The analysis employs events with 
reconstructed energy in the range 1--100~GeV, where the effective area of the 
LAT is largest and not strongly dependent on energy.
To allow the best separation between point sources and the 
structured interstellar emission in the analysis 
procedure, only front-converting events are used.

Events are selected from approximately 62~months of data from 
2008-08-11 until 2013-10-15 using the  standard low-residual 
CR background ``Clean'' events from the Pass~7 event selections
\footnote{The reprocessed data and instrument response functions P7REP\_CLEAN\_V15 are employed.}.
Exposure maps and the PSF for the pointing
history of the observations were generated using the standard
\fermilat{} ScienceTools package (version 09-34-02) available 
from the \fermi{} Science
Support Center\footnote{http://fermi.gsfc.nasa.gov/ssc/data/analysis/}.

\subsection{Fore-/Background Modelling}
Specialised IEMs are constructed to estimate the fore-/background interstellar
emission toward and through the $15^\circ \times 15^\circ$ region about the
GC.
The results of the \fermilat\ team study 
\cite{2012ApJ...750....3A}, which compared \GP-generated IEMs normalised to 
local CR measurements with \gray{} data, are used.
A major uncertainty affecting predictions of the interstellar 
emission toward the inner Galaxy is the spatial distribution of CR sources.
The pulsar distribution (``Pulsars'' \cite{2004A&A...422..545Y}) 
and the distribution of OB stars (``OBstars'' \cite{2000A&A...358..521B}) 
employed in the aforementioned \fermilat\ study encapsulate
this because they represent reasonable extremes for the Galactocentric radial 
dependence\footnote{The reader is referred to 
Fig.~1 in \cite{2012ApJ...750....3A} for a visualisation of the Galactocentric
radial variation.}.
These CR source distributions are used\footnote{
The models assume an axisymmetric cylindrical geometry for the CR confinement 
volume with a halo height $z_h = 6$~kpc and maximum radial 
boundary $R_h = 30$~kpc, and all other parameters the same so that the only
difference is the assumed spatial distribution for the CR sources.}
to calculate ``baseline'' \gray{} intensity templates for Galactocentric 
radial annuli for the standard 
processes ($\pi^0$-decay, IC, Bremsstrahlung\footnote{Bremsstrahlung is a
minor component over the 1--100~GeV energy range used in the analysis, and is 
held constant at the \GP\ predictions for each IEM.}).
The annular intensity maps are used as templates 
together with an isotropic component and a model 
for \gray{} emission associated with the Loop~I supernova remnant (SNR) 
employing a two-component spatial 
template from \cite{2007ApJ...664..349W} with a power-law spectral model 
for each, and 
point sources from the Third Fermi source catalogue 
(3FGL; \cite{2015ApJS..218...23A}).
The combined model for each IEM is fit to 
the \fermilat\ data {\it excluding the $15^\circ \times 15^\circ$ region about
the GC} using a maximum-likelihood method.

\begin{figure*}[ht]
\begin{overpic}[scale=0.55]{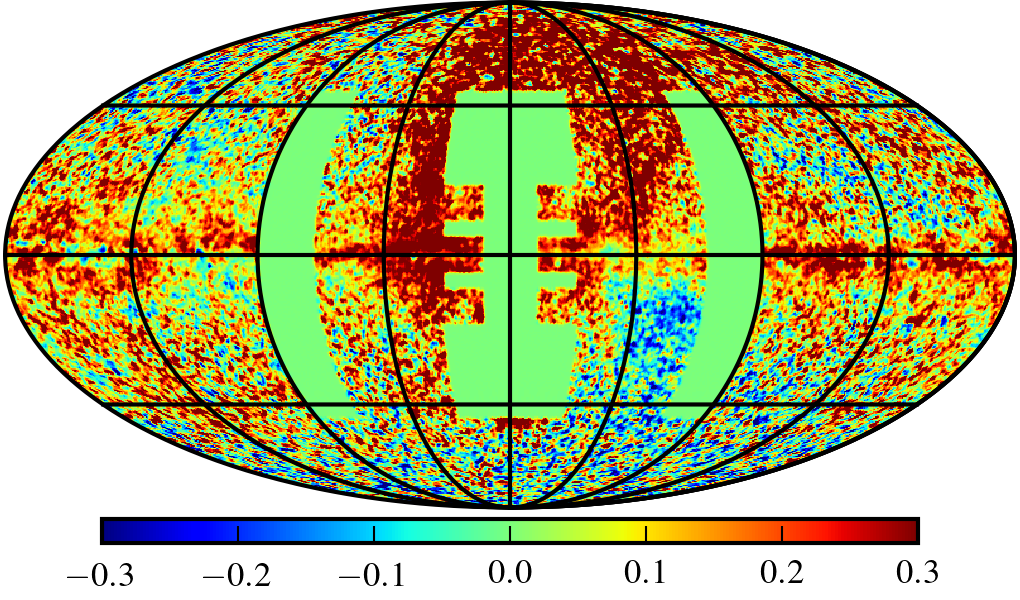} \put(32,60){Preliminary} \end{overpic}
\begin{overpic}[scale=0.55]{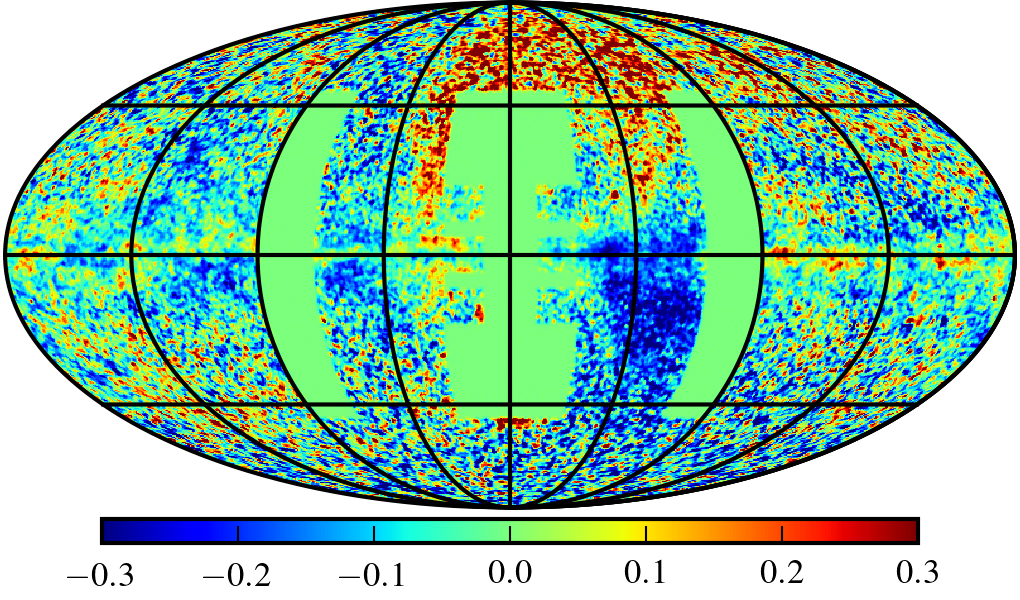} \put(32,60){Preliminary} \end{overpic}
\begin{overpic}[scale=0.55]{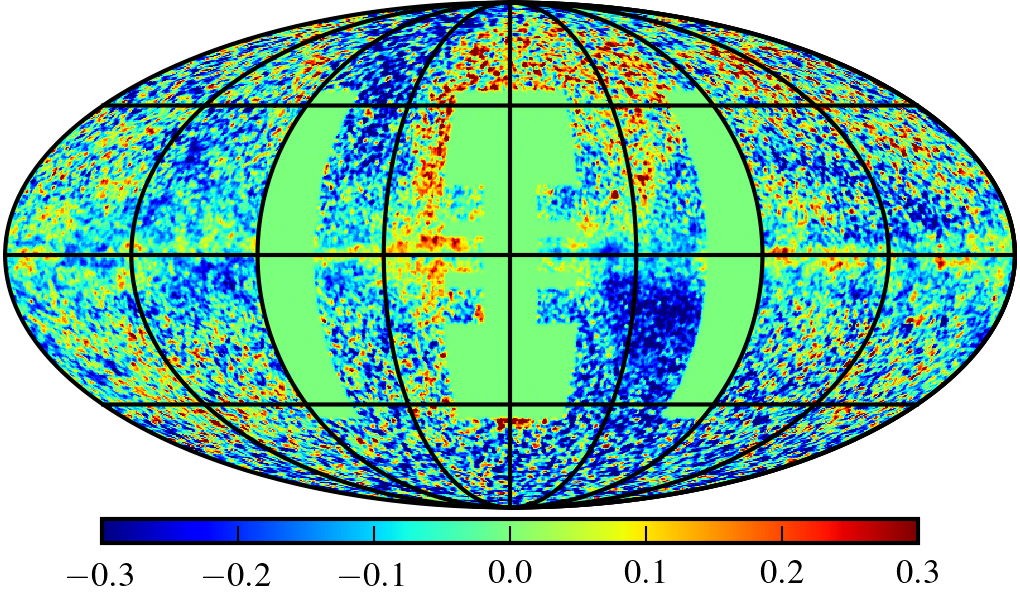} \put(32,60){Preliminary} \end{overpic}
\caption{Residual fractional counts $(data - model)/model$ in the 1--3.16~GeV
energy range for the baseline Pulsars model (left), 
intensity-scaled Pulsars model (centre), 
and index-scaled Pulsars model (right) fitted following the procedure 
described in the text.
The baseline model does not include a model for the Loop-I SNR, resulting 
in the positive residuals in the northern Galactic hemisphere inside the
solar circle.
The maps are calculated for a HEALPix \cite{2005ApJ...622..759G} 
order 8 pixelisation 
($\sim 0.23^\circ$ resolution) and smoothed with a 
1$^\circ$ FWHM Gaussian.
The positive residual at mid-to-high latitudes interior to the solar circle
is due to mismatch between the data and the relatively simple Loop~I model.
The residuals close to the plane caused by this simple model are lower and 
do not affect the analysis of the $15^\circ \times 15^\circ$ 
region about the GC.
\label{fig:residual_IEM}}
\end{figure*}

Figure~\ref{fig:residual_IEM} shows the fractional 
residuals, $(data - model)/model$, for the Pulsars IEM for the 
1--3.16~GeV energy band for the (left to right) baseline, and two IEM 
variants of this model that have been scaled to the data (described below).
The regions not used in the fitting procedure are explicitly masked in the
figure.
They are not used because of localised extended excesses that are most 
likely unrelated to the large-scale interstellar emission.

%\begin{overpic}[scale=0.55,grid,tics=10]{LRYusifovXCO5z6R30_Ts150_mag2_base_Residual_Fractional_Masked_1000-3162MeV.png}
%\put (32,60) {Preliminary}
%\end{overpic}

Outside of the Galactic plane the fractional residuals are substantially 
reduced for the IEM shown in the central panel compared to the baseline.
This IEM variant is termed `intensity-scaled' because only the 
normalisations of 
the individual intensity maps are adjusted in the scaling procedure.
The lower residuals for this IEM are due to the 
scaling of the $\pi^0$-decay interstellar emission for the
local annulus, and of the IC component generally.
Along the Galactic plane the $\gtrsim 30$\% under-prediction
by the baseline model is reduced to $\lesssim \pm10$\% after this scaling, 
except for scattered regions.
However, at higher energies (not shown) the intensity-scaled IEM tends
to over-predict the data along the plane interior to the solar circle.
To account for this, another variant is developed that includes additional 
degrees of freedom to the spectrum of the $\pi^0$-decay 
emission for annuli iterior to the solar circle. 
It is fit to the data following the same procedure as the 
intensity-scaled IEM.
This variant is termed `index-scaled' and has a fractional residual 
$\lesssim \pm 10$\% for longitudes $l \sim -(15-70)^\circ$ for the 
1--3.16~GeV band (Fig.~\ref{fig:residual_IEM}, right panel), with a 
slight increase in the residual for the corresponding positive longitude range;
the residuals improve to a similar degree at higher energies.
Qualitatively, similar results for the scaled OBstars 
IEMs (intensity/index-scaled) are also obtained.
It is not straightforward to identify a best IEM after fitting because the 
qualitative improvement for each over the corresponding 
baseline IEM is similar.
Consequently all 4 IEMs are used to estimate the fore-/background below.

\subsection{Modelling $15^\circ \times 15^\circ$ Region about the Galactic Centre}
\label{sec:ML}

Modelling of the interstellar emission and point sources over 
the $15^\circ \times 15^\circ$ region about the GC is accomplished 
using an iterative procedure that identifies point-candidates (`seeds') 
and fits for their fluxes and spectra together with the 
interstellar emission from $\pi^0$-decay associated with the neutral gas and 
IC components for the innermost annulus using a maximum-likelihood method, 
while the fore-/background interstellar emission (determined above) 
outside the inner $\sim 1$~kpc is held constant.

Point-source seeds are identified using 
the wavelet analysis algorithm 
{\it PGWave} \cite{1997ApJ...483..350D,Ciprini:2007zz}.
Application of {\it PGWave} identifies true point sources, as well as 
structures in the interstellar emission that are indistinguishable from 
point sources due to the 
finite angular resolution and statistics of the \fermilat\ data, without 
dependence on the specifics of an IEM. 
The spectra of candidates are initially evaluated using {\it PointLike}, a 
package for maximum-likelihood analysis of \fermilat\ 
data \cite{2011arXiv1101.6072K,2012ApJ...756....5L}.
This package is also used to optimise the positions of seeds from the 
{\it PGWave}-determined list.
%We use two hypotheses for the point-source spectra: a simple 
%power law, or a log-parabola model.
%The log-parabola model is used for a seed 
%if the {\it PointLike}-determined test 
%statistic ({\it TS}) for it is $>9$ compared to using the power-law model. 
%The source candidate selection procedure gives 67 seeds 
%with a {\it PointLike}-determined {\it TS}$>9$.
For the only extended source that has been previously identified in the 
region, the W28 supernova remnant \cite{2010ApJ...718..348A}, the 
spatial template and spectral model employed for the 3FGL analysis are used.
Its spectral parameters are refit during the maximum-likelihood procedure.

Point-source candidates are combined with their {\it PointLike} trial 
spectra together with the fore-/background models in a second 
maximum-likelihood fit. 
A binned likelihood 
fit is performed using the {\it Fermi} ScienceTool {\it gtlike}.
The templates for the $\pi^0$-decay related \gray{} intensity from \hi\ and 
CO, and the IC emission, in annulus~1 are freely scaled in the 
fitting procedure.
The point-source seed detection is run again on the residual maps 
to find fainter sources that were missed in the first iteration.
{\it PointLike} is again used to determine their initial spectra and optimise 
their localisations.
The combined point-source candidate list from the first and second iterations
\footnote{No new significant excesses that could be identified as viable 
seeds were found after two iterations of the procedure.}, and the 
interstellar emission components for annulus~1 are fit using {\it gtlike}.
The results of the maximum-likelihood fit are values and confidence ranges 
for the coefficients of 
the \hi\ annulus~1, CO annulus~1, IC annulus~1, as well as the
{\it TS}, fluxes and spectra for the point sources.
All point sources with a maximum-likelihood determined 
$TS > 9$ are included in the model.

\section{Results}

\subsection{Interstellar Emission}

\begin{figure*}[ht]
\begin{overpic}[scale=0.25]{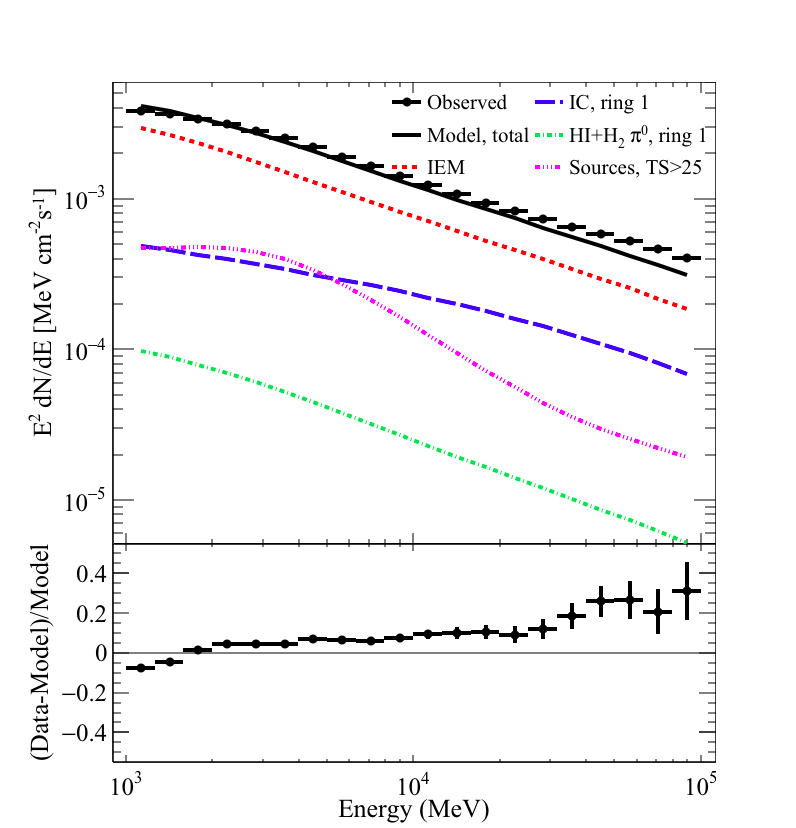} \put(40,92){Preliminary} \end{overpic}
\begin{overpic}[scale=0.25]{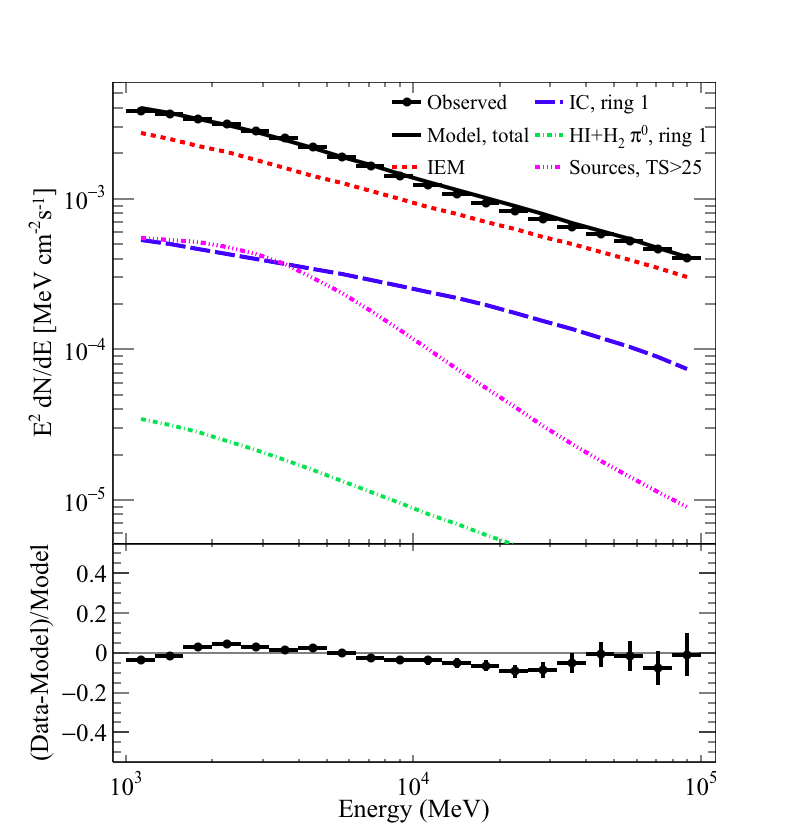} \put(40,92){Preliminary} \end{overpic}
\caption{
Differential flux components obtained for the Pulsars intensity-scaled
(left) and index-scaled (right) IEMs
for the $15^\circ \times 15^\circ$ region about the GC.
Line styles, as in figure legend.
%: solid (total model), long-dash (IC, annulus 1), 
%dot-dash (\hi\ and CO gas $\pi^0$-decay, annulus 1), 
%dot-dot-dot-dash (point sources), 
%dash (Galactic interstellar emission excluding annulus 1 for IC, \hi\ and 
%CO gas $\pi^0$-decay). 
Solid circles: data.
\label{results:interstellar_emission_fluxes}}
\end{figure*}

Figure~\ref{results:interstellar_emission_fluxes} shows the differential 
spectra of the individual components obtained for the Pulsars 
intensity-scaled and index-scaled IEMs
integrated over the $15^\circ \times 15^\circ$ region about the GC 
(the results for the OB stars IEMs are similar, but not shown due to
space limitations).
The figure separates the emission components in terms of the contributions
by $\pi^0$-decay and IC for annulus~1, the interstellar emission 
fore-/background,
and point sources over the region.
As expected, the fore-/background dominates for each IEM, which is
predominantly $\pi^0$-decay in origin.
Interestingly, 
the IC is the dominant interstellar emission component over the 
inner $\sim1.5$~kpc, and is much higher than predicted by \GP\ for the
baseline IEMs.
Its contribution to the total flux depends on the IEM and whether the residual
is fitted (Section~\ref{results:residuals}).
The variation of the IC flux for annulus~1 over all 4 IEMs is 
within a factor $\sim 1.5$. 
The interstellar emission from the neutral gas $\pi^0$-decay is considerably
suppressed compared to the \GP\ predictions.
For both Pulsars and OBstars models 
the interstellar emission fore-/background is harder at high
energies for the index-scaled IEMs, which reduces the spectral residuals 
for energies $\gtrsim 10$~GeV.

\subsection{Point Sources}

\begin{figure*}[ht]
\begin{overpic}[scale=0.7]{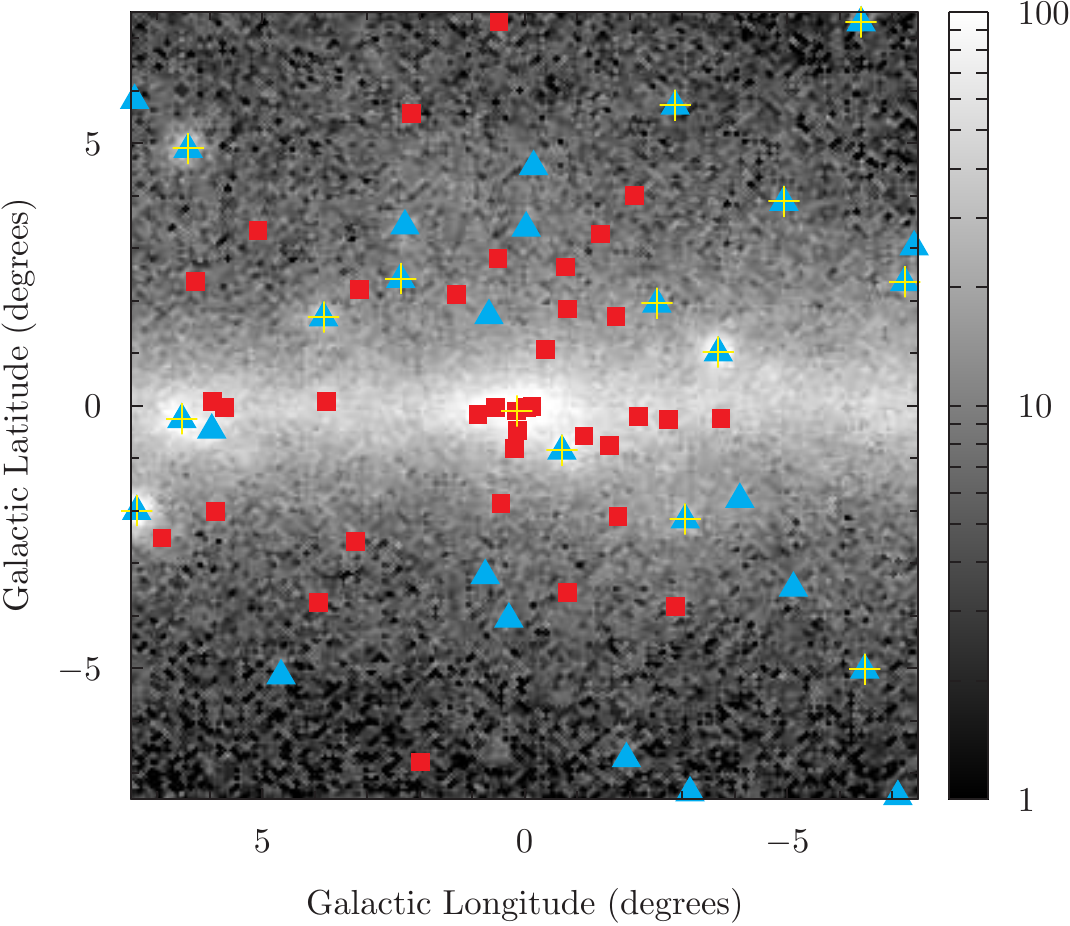}\put(40,87){Preliminary} \end{overpic}
\begin{overpic}[scale=0.7]{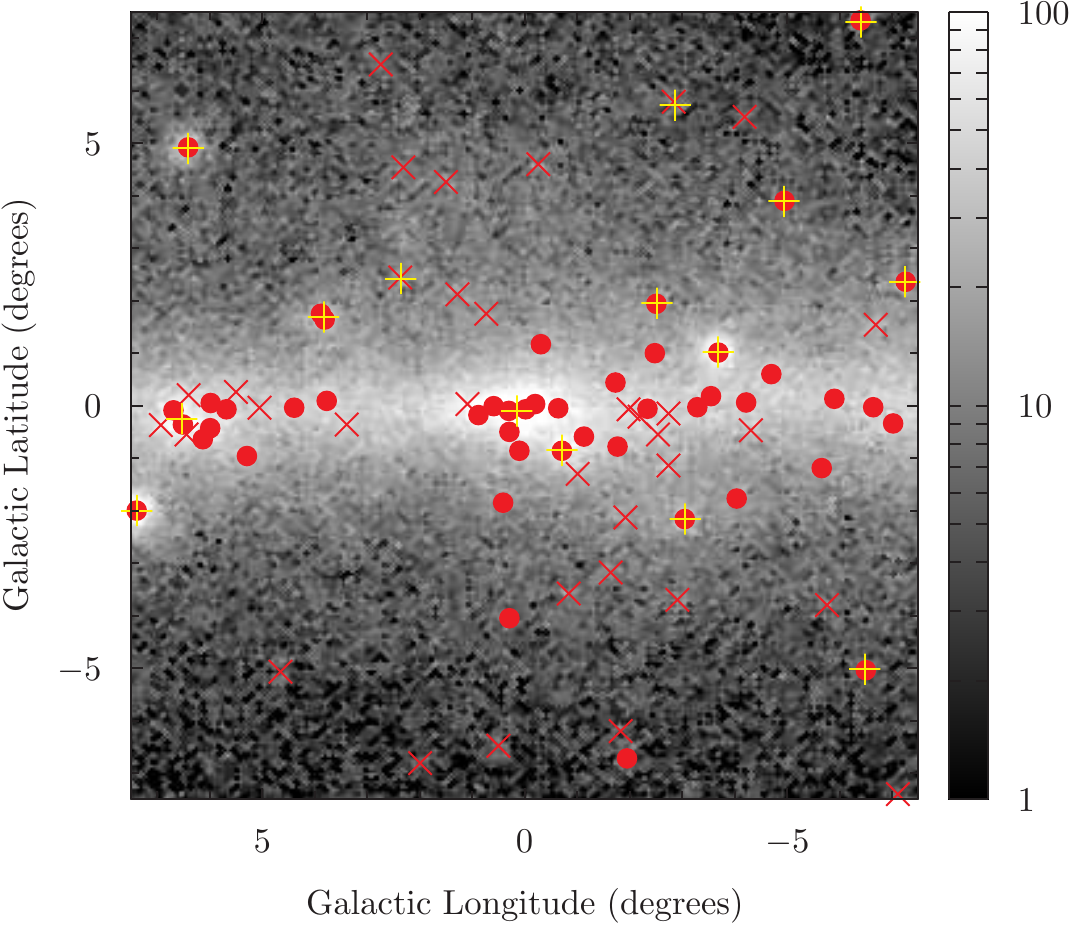}\put(40,87){Preliminary} \end{overpic}
\caption{
Point sources for 3FGL (left panel) and 1FIG (right panel, for Pulsars 
intensity-scaled IEM) overlaid on 
the total counts for the $15^\circ \times 15^\circ$ region about the GC.
Left panel symbol key: filled squares, `flagged' 3FGL sources; 
filled triangles, other 3FGL sources; upright crosses, 3FGL sources with a 
multi-wavelength association. 
Right panel symbol key: filled circles, 1FIG sources with $TS \geq 25$; 
angled crosses, 1FIG source candidates with $TS < 25$; upright crosses, 
as in left panel.
Colour scale is in counts per $0.05^2$ degree pixel.
\label{fig:3fgl_1fig_overlay}}
\end{figure*}

The analysis finds 48 point sources over the $15^\circ \times 15^\circ$ region 
about the GC with a $TS \geq 25$ for the Pulsars intensity-scaled IEM,
which was the model used for the point-source positions and localisation 
uncertainties\footnote{The optimisation of the source seed localisation does not have a strong dependence on the IEM.}.
This collection is denoted the 1$^{\rm st}$ Fermi Inner
Galaxy (1FIG) source list.
Figure~\ref{fig:3fgl_1fig_overlay} shows the point sources from the 3FGL and 
1FIG overlaid on the total photon counts for the $15^\circ \times 15^\circ$
region about the GC.
The 3FGL sources are separated
according to whether they have an analysis flag set in the 3FGL catalogue: 
flagged sources indicate their properties depend on the IEM or other details
of the analysis in the region.
The density of flagged 3FGL sources is higher out of the Galactic plane than 
that of the 1FIG sources, even if the $TS < 25$ source candidates are included.
This can be partly attributed to differences in the IEMs used for the 
respective analyses.

Some of the 1FIG sources and source candidates are likely due to mismodelling
of the interstellar emission.
However, it is difficult to determine the precise fraction.
Multi-wavelength association with objects of known 
\gray{} emitting source classes can suggest counterparts.
There are 14 1FIG source and source candidate overlaps when comparing 
with SNRs from Green's SNR 
catalogue\footnote{http://www.mrao.cam.ac.uk/surveys/snrs/} \cite{2014BASI...42...47G}
and pulsars from the ATNF catalogue\footnote{http://www.atnf.csiro.au/people/pulsar/psrcat/} \cite{2005AJ....129.1993M}. 
Multiple overlaps occur across and within the catalogues, e.g., 
SNR~354.1+00.1 and PSR~J1701-3006A,D,E overlap with 1FIG~J1701.1-3004.
Three of the 1FIG sources overlap with SNRs that have not previously been
detected in high-energy \gray{s}; 
follow-on studies are required to better characterise their spatial and 
spectral properties.

\subsection{Residuals}
\label{results:residuals}

\begin{figure*}[htb]
\begin{overpic}[scale=0.5]{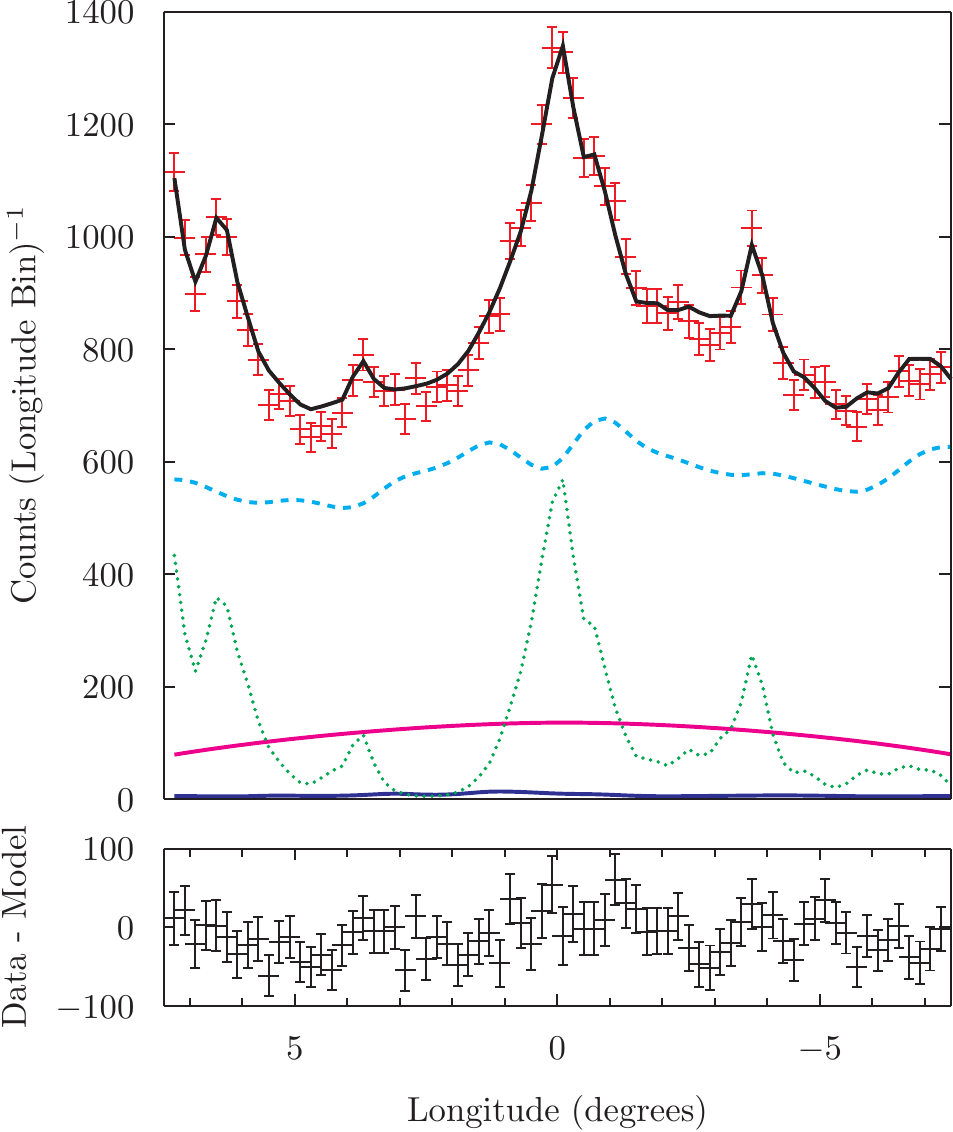}\put(35,102){Preliminary} \end{overpic}
\begin{overpic}[scale=0.5]{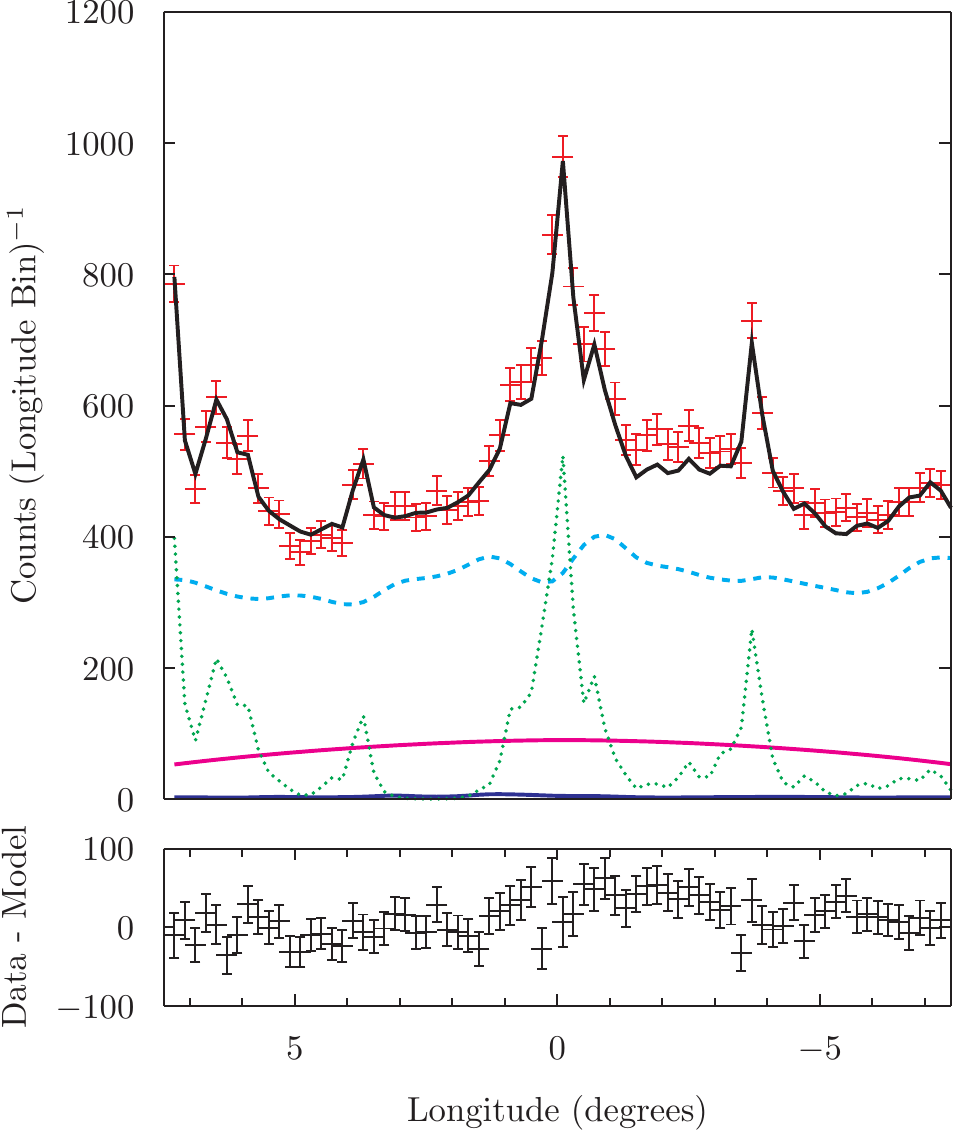}\put(35,102){Preliminary} \end{overpic}
\begin{overpic}[scale=0.5]{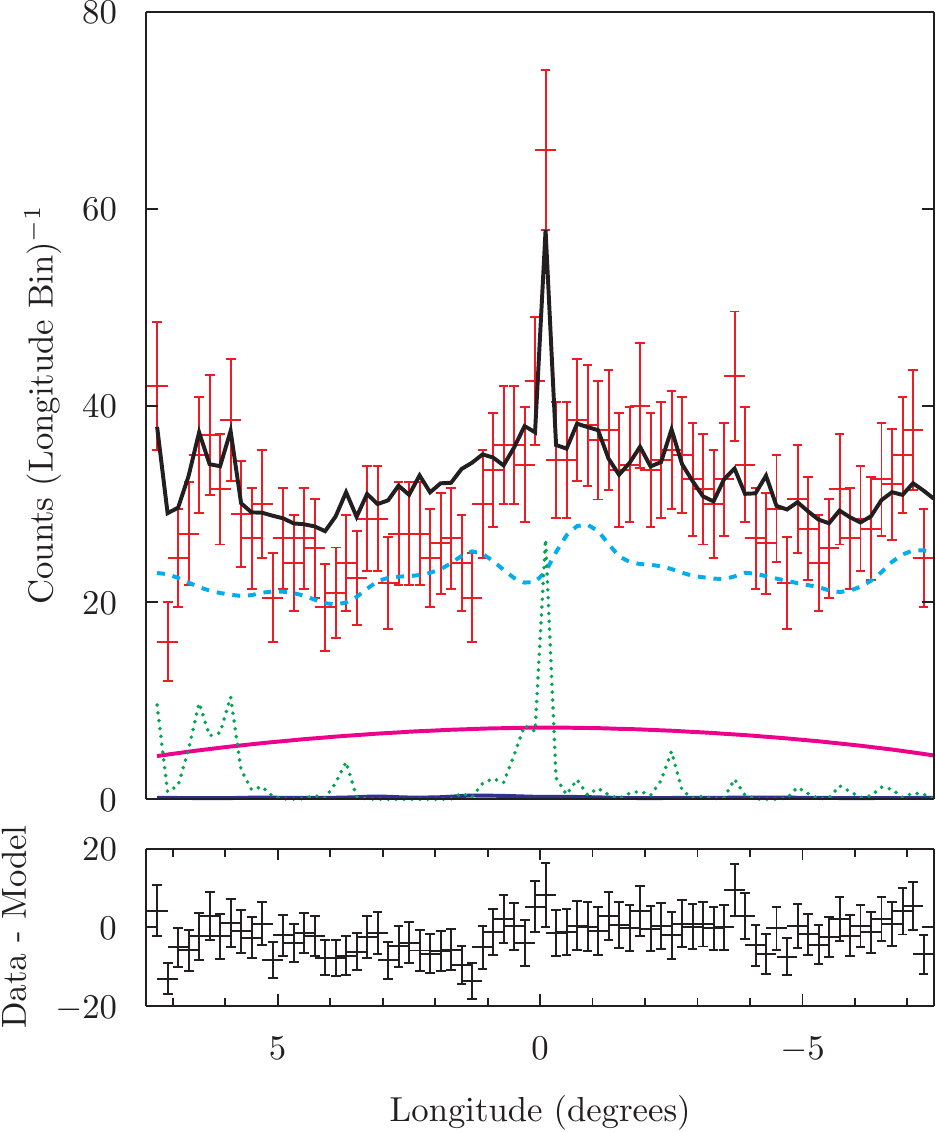}\put(35,102){Preliminary} \end{overpic}
\caption{
\label{fig:res_profiles}
Longitude profiles for 1--1.6~(left), 1.6--10~(middle), and 10--100~GeV 
energies (left), respectively, of the residual 
counts, $data-model$, for the Pulsars 
index-scaled IEM after fitting for interstellar emission and point sources 
across the $15^\circ \times 15^\circ$ region.
Line styles: black/solid, total model; cyan/dashed, fore/background interstellar
emission; green/dotted, point sources; magenta/solid, IC from annulus~1; 
blue/solid, $\pi^0$-decay from annulus~1.
Point styles: red, data; black, residual counts.
The lower sub-panel for each profile gives the residual counts after the model
has been subtracted from the data.
The error bars are statistical.
Profiles for the residuals counts for other IEMs display similar features 
with the major difference being the number of counts.}
\end{figure*}

Figure~\ref{fig:res_profiles} shows the longitude profiles for 
the 1--1.6, 1.6--10, and $>10$~GeV energy ranges\footnote{Each band
covers the energy intervals where the under/over-predictions 
for the fractional residuals are more prominent in Fig.~\ref{results:interstellar_emission_fluxes}.} for the 
Pulsars index-scaled model, which is the model that
has the lowest fractional residual across the 1--100~GeV energy range .
The lower sub-panel for each figure gives the residual counts, $data-model$.
(The features are mostly the same for the other IEMs.)
% with the major 
%difference being their magnitude in counts. Therefore, the profiles are 
%not shown because of their similarity.)
While there is considerable statistical noise, 
there is some indication that the residual counts are distributed 
asymmetrically in longitude about the GC below 10~GeV.
%There is also a weak indication of anti-correlation between the 1--1.6~GeV and 
%1.6--10~GeV residual counts.

The model over prediction at the lowest energies is 
primarily correlated 
with the Galactic plane, which could be due to mismodelling of the 
gas component of the IEMs. 
Some of the positive residual in the few GeV range could be due to an
extended component that is more concentrated toward the GC compared to the 
IEM components.
However, it is difficult to establish properties for an additional
component not presently included in the model for the region.
%A needs to be assumed and fit together with 
%the interstellar emission and point source contributions (Sec.~\ref{sec:ML}).
Different spatial models are considered for the positive residuals: a 
squared Navarro-Frenk-White (NFW) profile \cite{1997ApJ...490..493N} for 
dark matter annihilation, 
2D Gaussians with HWHM $1^\circ$, $2^\circ$, $5^\circ$, and $10^\circ$, and 
the CO annulus~1 gas template smoothed with a $2^\circ$ Gaussian as a tracer
of unresolved pulsars.
An exponential cutoff power law is used for the spectral model, which 
has some flexibility to model a pulsar or a DM annihilation 
spectrum without supposing specific scenarios.
The positive residual model is fit together with the interstellar 
emission and point sources (Sec.~\ref{sec:ML}). 
Its flux and spectral parameters depend
%Tflux and spectral parameters for the positive residual obtained 
%from fitting these models depend 
strongly on the interstellar emission fore-/background, and 
%The positive residual model parameters 
are also covariant with
the normalisations of the standard interstellar emission 
processes ($\pi^0$-decay and IC) for annulus~1.
% that are also fit over the 
%$15^\circ \times 15^\circ$ region.
Note that not all of the positive residual is accounted for by the centrally 
peaked models that are considered here.
% in this analysis.
The remainder is distributed about the Galactic plane, but with no obvious 
correlation to the spatial templates used for the analysis.
%As a consequence it is not straightforward to identify a singular origin for it.
% (Section~\ref{sec:ML}).

\acknowledgments
The \textit{Fermi}-LAT Collaboration acknowledges support for LAT 
development, operation and data analysis from NASA and DOE (United States), 
CEA/Irfu and IN2P3/CNRS (France), ASI and INFN (Italy), MEXT, KEK, and 
JAXA (Japan), and the K.A.~Wallenberg Foundation, the Swedish Research 
Council and the National Space Board (Sweden). Science analysis support in 
the operations phase from INAF (Italy) and CNES (France) is also gratefully 
acknowledged.

GALPROP development is partially funded via NASA grants NNX10AE78G and 
NNX13AC47G.

SM's work is supported by a fellowship from the Hellman Foundation and the US DOE.

%Some of the results in this paper have been derived using the 
%HEALPix~\cite{2005ApJ...622..759G} package.

\end{document}